\documentclass[a4paper, amsfonts, amssymb, amsmath, reprint, showkeys, nofootinbib]{revtex4-1}
\usepackage[english]{babel}
\usepackage[utf8]{inputenc}
\usepackage[colorinlistoftodos, color=green!40, prependcaption]{todonotes}
\usepackage{amsthm}
\usepackage{mathtools}
\usepackage{physics}
\usepackage{xcolor}
\usepackage{graphicx}
\usepackage[left=19mm,right=16mm,top=20mm,columnsep=15pt]{geometry} 
\usepackage{adjustbox}
\usepackage{placeins}
\usepackage[T1]{fontenc}
\usepackage{lipsum}
\usepackage{csquotes}
\usepackage[pdftex, pdftitle={Article}, pdfauthor={Author}]{hyperref} % For hyperlinks in the PDF

\usepackage{hyperref}
\begin{document}
\title{Scanning a focus through scattering media without using the optical memory effect}

\author{Bahareh Mastiani}
\email[Corresponding author: ]{b.mastiani@utwente.nl}
\affiliation{Biomedical Photonic Imaging Group, Faculty of Science and Technology, University of Twente, P.O. Box 217, 7500 AE Enschede, The Netherlands}

\author{Tzu-Lun Ohn}
\affiliation{Biomedical Photonic Imaging Group, Faculty of Science and Technology, University of Twente, P.O. Box 217, 7500 AE Enschede, The Netherlands}

\author{Ivo M. Vellekoop}
\affiliation{Biomedical Photonic Imaging Group, Faculty of Science and Technology, University of Twente, P.O. Box 217, 7500 AE Enschede, The Netherlands}

\date{\today} 

\begin{abstract}
Wavefront shaping makes it possible to form a focus through opaque scattering materials. In some cases, this focus may be scanned over a small distance using the optical memory effect. However, in many cases of interest, the optical memory effect has a limited range or is even too small to be measured. In such cases, one often resorts to measuring the full transmission matrix (TM) of the sample to completely control the light transmission. However, this process is time-consuming and may not always be possible. We introduce a new method for focusing and scanning the focus at any arbitrary position behind the medium by measuring only a subset of the transmission matrix, called Sparse Field Focusing (SFF). With SFF, the scan range is not limited to the memory effect and there is no need to measure the full transmission matrix. Our experimental results agree well with our theoretical model. We expect this method will find applications in imaging through scattering media, especially when the optical memory effect range is small.
\end{abstract}

\maketitle
It is challenging to perform high resolution imaging deep inside scattering media. Due to the inhomogeneity of the refractive index, light is scattered during propagation. However, wavefront shaping techniques can compensate for this distortion and achieve a focus despite scattering \cite{vellekoop2007focusing, kubby2019wavefront}. For imaging applications, it is desirable to scan the constructed focus in a two-dimensional plane behind or inside a scattering medium.

\begin{figure}
\centering
\includegraphics[width=\linewidth]{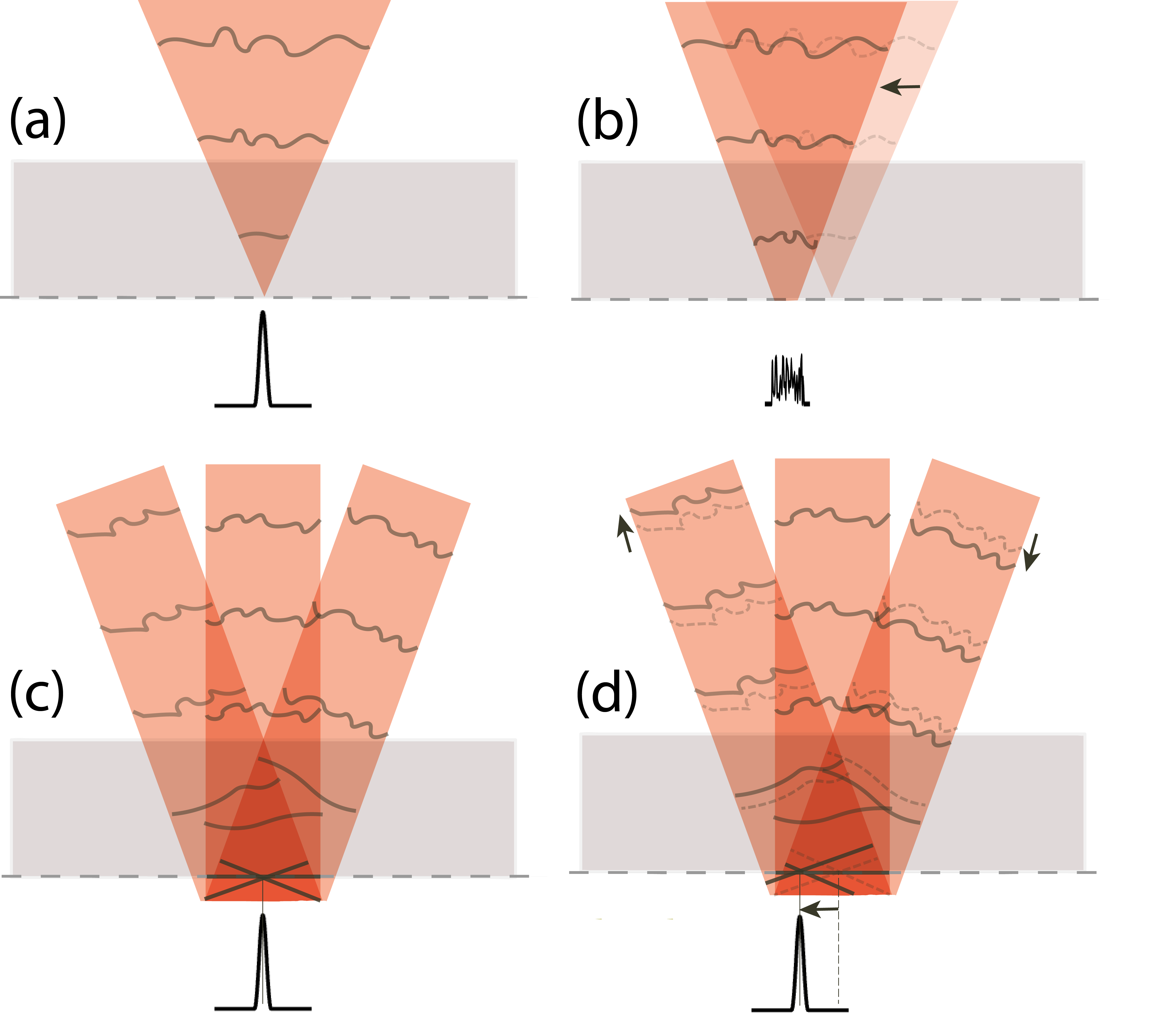}
\caption{Focusing and scanning via conventional methods (a, b) and SFF (c, d). (a) Focusing scattered light through a turbid medium. (b) Shifting the incident wavefront destroys the  focus. (c) A discrete set of shaped waves illuminates the medium. The focus is formed by superposition of the resulting plane waves in the image plane. (d) Changing the relative phase (solid line) between wavefronts makes the focus move. Note: for clarity, only three non-overlapping incident waves are drawn. In the experiment, several hundreds of overlapping incident waves are used.}
\label{fig:concept}
\end{figure}

There are two approaches to scan the focus acquired by wavefront shaping. First, a single corrected focus is scanned to a new position by applying a shift and/or tilt to the incident wavefront. When the correlation of corrections between the new and the previous positions is high, the previous correction can also be applied to form a focus in the new position. The correlation is called the optical memory effect. \cite{freund1988memory, judkewitz2015translation, Schott:15, osnabrugge2017generalized, vellekoop2010scattered, park2015high}. In this approach, a focus can be scanned over a limited distance called the memory-effect range or, alternatively, isoplanatic patch. However, to focus light at a position outside of this range, a different correction is needed. This problem is illustrated in Fig.~\ref{fig:concept}(a, b). Figure.~\ref{fig:concept}(a) shows the required shaped wavefront to focus the scattered light through a strongly scattering layer. In Fig.~\ref{fig:concept}(b), shifting the incident wavefront causes the light to propagate through a different part of the sample than the wavefront was originally constructed for. Therefore, the shifted wavefront will not form a focus. This is the limitation of the ordinary scanning technique for scanning a focus. The optical memory-effect range is small when focusing deep inside biological tissues(e.g. 6 $\mu m$ through 1 $mm$ of chicken breast tissue) \cite{judkewitz2015translation}. 

The second approach is to measure the full transmission matrix of the scattering medium. The transmission matrix (TM) is the input-output response of the scattering medium. Knowledge of the full TM makes it possible to focus at any arbitrary position behind the medium \cite{popoff2010measuring, yu2013measuring, kim2015transmission}. Furthermore, the measured TM provides the required information to transmit images through the medium \cite{popoff2010image, mosk2012controlling}. However, measuring the full TM, if at all possible, is computationally expensive, memory consuming, and slow.

In this paper, we present a method in which a corrected focus is scanned through a strongly scattering medium beyond its isoplanatic patch, by only measuring a subset of the transmission matrix.
We will first present our new concept of Sparse Field Focusing (SFF), and experimentally demonstrate that it achieves a scan range that far exceeds the isoplanatic patch. Afterwards, we present an analytical model for SFF and compare it to our experimental results.

In our method, the scattering medium is illuminated by a group of superposed wavefronts. Each of these wavefronts is shaped (optimized) in such a way that, after scattering, the light forms a plane wave at the back side of the sample (the image plane). Superposing the optimized wavefronts creates a focus as a result of constructive interference of the transmitted plane waves at the image plane.

Figure.~\ref{fig:concept}(c, d) shows the principle of SFF. In Fig.~\ref{fig:concept}(c)  the scattering medium is illuminated by a set of wavefronts optimized to form apodized plane waves at the image plane. The apodized plane waves interfere constructively and form a focus at the image plane. Mathematically, the field in the image plane, ${E}(x,y)$, is now a linear combination of $M$ apodized plane waves added to the background field

\begin{equation}
{E}(x,y) = \sum_{m=1}^{M}A_m(x,y)\exp{(ik_{mx}x+ik_{my}y + i\varphi_m)}+E_{bg},
\label{eq:superposition}
\end{equation}

where $A_m(x,y)$ is the amplitude of the $m^{th}$ optimized field in the image plane. $k_{mx}$ and $k_{my}$ are the components of the wave vector parallel to the image plane, and $\varphi_m$ is an optional and additional phase shift. When $\varphi_m=0$, the waves interfere constructively to form a focus at $x=0,y=0$. Practically, not all the light can contribute to form plane waves, a part of the light forms a background speckle, $E_{bg}$. (see Appendix A for the quantitative treatment).

Figure~\ref{fig:concept}(d) shows the proposed method of scanning the focus constructed using SFF. To scan the focus by a distance of $\Delta x$ and $\Delta y$, we apply phase shifts of $\varphi_m=k_{mx}\Delta x+k_{my}\Delta y$ to the incident waves. Since we only change the overall phase of each shaped wave, each individual wave will still propagate through the medium in the exact same way as without phase shift. From \eqref{eq:superposition}, we see that the waves now interfere constructively at a shifted position $x=\Delta x$ and $y=\Delta y$. In other words, the focus has been shifted without the need to use a wavefront-shaping algorithm to re-calculate the wavefront for focusing in new isoplanatic patches.

\begin{figure}
\centering
\includegraphics[width=\linewidth]{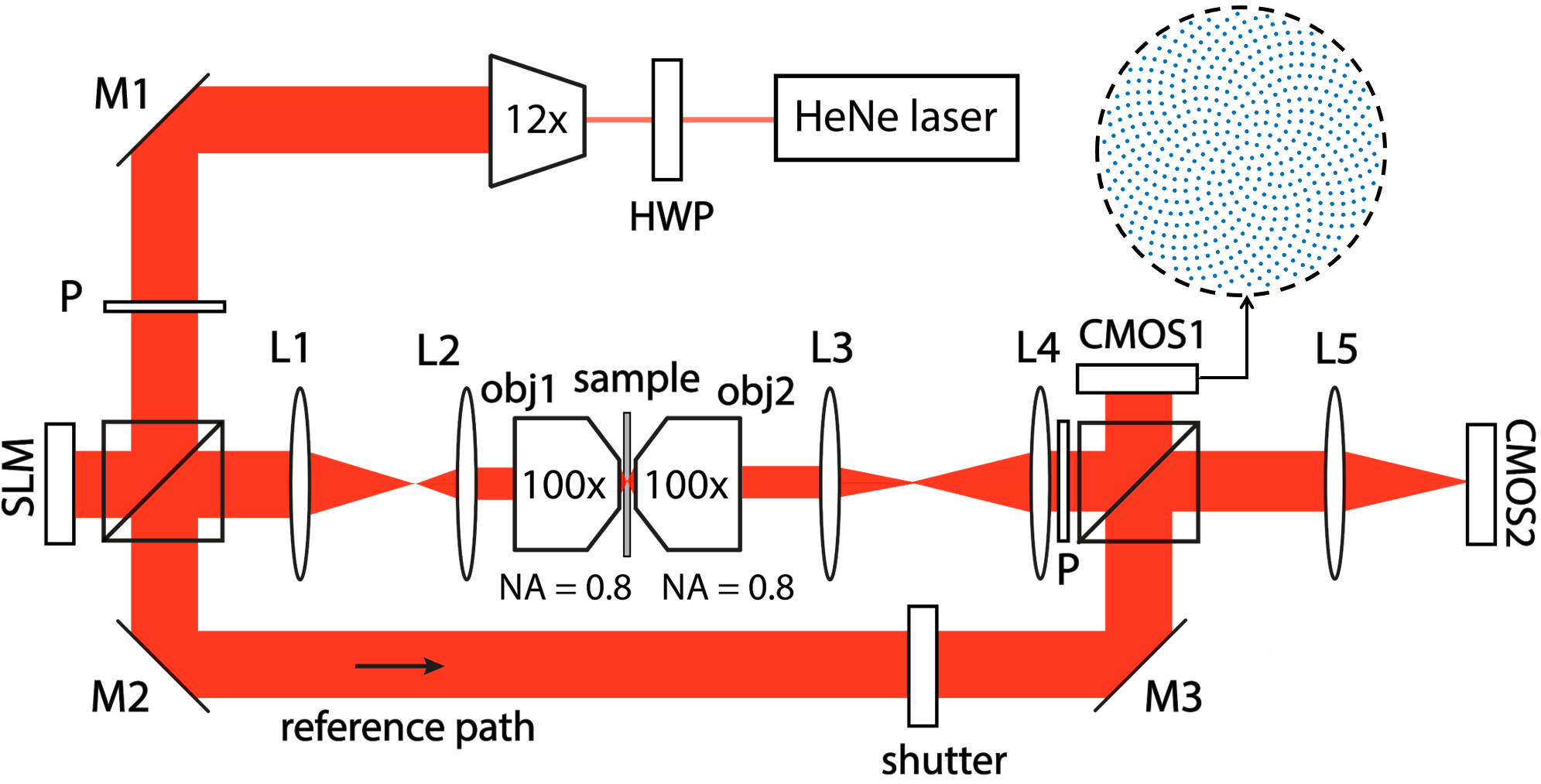}
\caption{Schematic of the experimental setup. HWP, half-wave plate; M, mirror; BS, 50\% non-polarizing beam splitter; P, polarizer; CMOS, complementary metal oxide semiconductor camera; obj, microscope objective lens; L1, L2, L3, L4 and L5, lenses with focal length of respectively 150 mm, 75 mm, 100 mm, 150 mm and 200 mm. The inset shows the distribution of 500 targets on CMOS1 and the circle around the targets corresponds to the back pupil of obj2.}
\label{fig:setup}
\end{figure}

We used the setup depicted in Fig.~\ref{fig:setup} to measure the partial TM of a scattering sample. A 632.8 nm HeNe laser beam was expanded and split into two paths, one was reflected off a phase-only spatial light modulator (GAEA-2 NIR, Holoeye). After a 4f system, the SLM was imaged onto the back focal plane of a microscope objective (A-Plan 100$x$/0.8, Zeiss), which focused the light onto the surface of the sample. The intensity distribution at the back focal plane of the second objective was imaged with the CMOS1 camera (acA2000-165umNIR, Basler), providing feedback for wavefront shaping. An identical CMOS camera (CMOS2) was used only to visually inspect the focus that is formed by SFF, it was not used for wavefront shaping.

The sample under study is a 11$\pm$3 $\mu m$ thick layer of zinc-oxide (Sigma Aldrich, average grain size 200 nm) on a coverslip with a thickness of 170 $\mu m$. The transport mean free path of similar zinc-oxide samples was measured to be around 0.6 $\mu m$ at a wavelength of $\lambda$ = 632.8 nm \cite{vanPutten}. Consequently, the sample is optically thick so that there is no transmitted ballistic light. Due to the isotropic scattering, there is no anisotropic memory effect \cite{judkewitz2015translation}, therefore this effect cannot be used to scan the focus inside (or at the back surface) of the sample.

 We performed wavefront shaping by running a stepwise sequential algorithm \cite{vellekoop2008phase}, and obtaining feedback from 500 individual targets on CMOS1. Each target is a circular region with a diameter of 3 pixels, corresponding to 11 $\mu m$ at the back pupil plane of obj2, which is smaller than the speckle size in that plane (diameter of 30 $\mu m$). The locations of the targets are shown in Fig.~\ref{fig:setup}(inset). During the wavefront shaping process, the reference path is blocked. After performing wavefront shaping, each optimized wavefront, corresponding to a row of the transmission matrix, provides the information needed to construct a focus in the back focal plane (CMOS1), and consequently a plane wave in the image plane. 
 
Next, we superposed these optimized waves in the image plane in order to make a focus. For this superposition, all the optimized waves need to be in phase to have constructive interference. The wavefront shaping algorithm does not give the overall phase for each wave, so an interferometric measurement is needed to determine the relative phase among the optimized waves. The optimized waves have the same phase as the original speckle pattern \cite{vellekoop2007focusing}, therefore it suffices to measure the phase of the original speckle pattern in the targets on CMOS1 using phase- step holography. We unblocked the reference path and performed phase-step holography. Once the relative overall phase among waves has been measured, we blocked the reference path and subtracted the optimized wavefronts to their measured overall phase so that all optimized waves are in phase. Finally, we summed the $M$ optimized fields in order to generate a superposition of $M$ optimized waves in the image plane. After displaying the phase of this superposed field on the SLM, a bright focus appeared in the image plane.
 
\begin{figure}
\centering
\includegraphics[width=\linewidth]{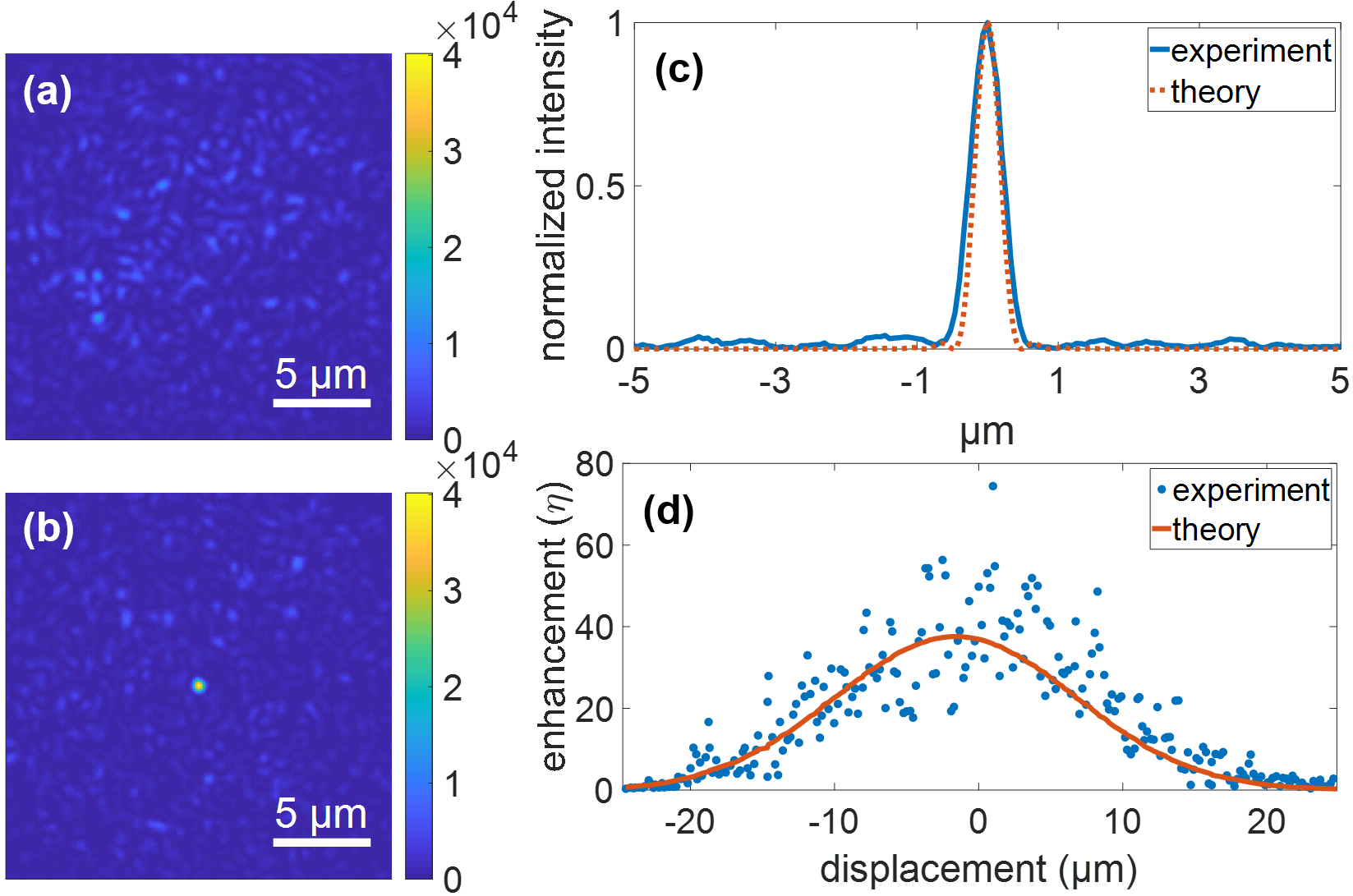}
\caption{(a) Intensity at the image plane with a non-shaped incident beam, and (b) with combining 500 optimized plane waves. The acquired focus is 47 times brighter than the original speckle pattern.  (c) Intensity profile of the formed focus at the image plane. Dashed line is the intensity profile of the theoretical diffraction-limited focus (d) Measured intensity enhancement as a function of the displacement from the center for the vertical scanning (blue circles). The predicted value for the enhancement, as given by Eq. \ref{eq:enhancement}, is represented by the red solid line.}
\label{fig:result}
\end{figure}

 Figure~\ref{fig:result}(a) shows the random speckle pattern in the image plane when an unshaped beam is focused onto the sample. In Fig.~\ref{fig:result}(b) the result of interfering 500 optimized waves is shown. A bright focus is formed in the image plane. Figure.~\ref{fig:result}(c) shows the intensity profile of the focus acquired by interfering 500 waves, and the intensity profile of the theoretical diffraction limited focus (dashed red). The acquired focus is 47 times brighter than the non-optimized speckle pattern. The full width at half maximum of the obtained focus and the theoretical diffraction-limited focus are 0.520 $\mu m$ and 0.407 $\mu m$, respectively. This slight difference is most likely caused by small misalignment of CMOS2, not perfectly placed in the plane conjugated to CMOS1, or the aberration caused by L5.

Next, we tried to test how far the focus can be scanned by manipulating the relative overall phase among the optimized waves. First, we calculated the required phase shift, $\theta$, for scanning the focus for a specified displacement along the vertical axis in the image plane. Next, to determine the phase shift of each incident wavefront, $\varphi_m$, (Eq. \ref{eq:superposition}), we linearly mapped the range from 0 to $\theta$ to the vertical coordinates of 500 targets position shown in Fig.~\ref{fig:setup}(inset). By applying the phase shift to the incident wavefronts, the focus was scanned along the vertical axis in the image plane (see Visualization 1). We measured the enhancement of the scanned focus, defined as the ratio of the optimized intensity at the focus location to the reference intensity which is the averaged intensity over 100 positions for the sample \cite{kubby2019wavefront}. In Fig.~\ref{fig:result}(d) the measured enhancement is plotted as a function of the spatial displacement of the focus from the beam center during the vertical scanning. The focus can be scanned over a range of 30 $\mu m$.

Finally, we verified that the formed focus could not be shifted using the optical memory effect applying tilt or shift or the combination of them. It only causes the intensity of the focus to decrease (data are not shown).

We present an analytical model that describes the enhancement of a focus coming from a superposition of $M$ optimized fields (Eq. \ref{eq:superposition}). In Appendix A, we show that the enhancement of this focus, $\eta(x,y)$, equals

%---
\begin{equation}
\eta(x,y) = \frac{|\gamma|^2M(N-1)}{N_s}F(x,y)+1,
\label{eq:enhancement}
\end{equation}
%---
where $M$ is the number of optimized fields, $N$ is the number of controlled segments on the SLM used for wavefront shaping. $|\gamma|^2$ is the wavefront shaping fidelity describing the quality of the wavefront modulation \cite{vellekoop2015feedback, yilmaz2013optimal} (see Appendix A). $N_s$ is the number of speckles in the background calculated by dividing the area of back pupil of objective to the area of one speckle. $F(x,y)$ is the distribution of the averaged intensity in the image plane over disorders of the sample, when the SLM is displaying the corrected wavefront. $F(x,y)$ is normalized to have a maximum of one. In our experiment, the average fidelity of the wavefront shaping was measured to be 0.54 with $N=1010$. The predicted enhancement for the scanned focus using Eq. (\ref{eq:enhancement}) is shown in Fig.~\ref{fig:result}(d) (red solid).

The focus is scanned for $n$ individual scanning points by changing the relative phase between $M$ incident wavefronts. In other words, we perform a sparse sampling of $M$ rows of the transmission matrix. Measuring only $M$ transmission matrix rows is sufficient to focus at $n$ positions through scattering media. To quantify how efficient SFF can correct in various isoplantic patches, the gain, $A$, is defined as the total number of individual scan points ($n$) divided by the number of shaped wavefronts ($M$), which is given by the equation:

\begin{equation}
A = \frac{n}{M}.
\label{eq:gain}
\end{equation}

In our experiment, two-dimensional scanning in the image plane provided 2984$\pm$390 individual scanning points with an enhancement higher than 10. Using Eq. (\ref{eq:gain}), the calculated gain for our experiment is 6.0$\pm$0.8 which means that we obtained the full transmission matrix by performing 6 times less measurement than measuring the full transmission matrix.

In Fig.~\ref{fig:result}(d) the measured enhancement decreases, where the focus is scanned further away from the center of the beam as the model (red solid) ,given by Eq. \ref{eq:enhancement}, predicts. This decrease follows the spatial distribution of the ensemble averaged intensity in the image plane, which results in a low enhancement for the focus positions further away from the beam center, and limits the scan range.

According to Eq. (\ref{eq:enhancement}) and Eq. (\ref{eq:gain}), there is a trade-off between the gain ($A$) and enhancement ($\eta$). To increase $\eta$, we can increase the number of optimized waves ($M$), but the number of measurement increases causing the gain to decrease.

In summary, we have presented a novel focusing and scanning through scattering media with fewer measurement than the conventional methods \cite{vellekoop2007focusing,popoff2010measuring,wang2015direct}. This method enables us to scan the focus approximately 30 $\mu m$ through a strongly scattering medium, while the isoplanatic patch is less than 0.38 $\mu m$. Our experimental results show that we can scan the focus through the scattering medium for a number of individual points which is 6.0$\pm$0.8 times greater than the number of the measured rows of the transmission matrix. The enhancement of the intensity of the focus can be described by Eq. (\ref{eq:enhancement}). We confirmed that we can achieve the similar resolution by measuring only part of the transmission matrix as when measuring the full transmission matrix. Furthermore, focusing inside scattering media has been done with an embedded guide star \cite{hsieh2010digital,vellekoop2008demixing,katz2014noninvasive,aulbach2012spatiotemporal}. Since it is not possible or desirable to have a guide-star everywhere, we also envision that our sparse sampling approach may be adapted to imaging inside scattering media using a subset of embedded guide-stars.

\section*{APPENDIX A} \label{section:AppendixA}
In this appendix, we derive the enhancement of the constructed focus using SFF (Eq. (\ref{eq:enhancement})). In our analytical model, we describe scattering in the sample with the transmission matrix elements, $t_{ba}$, which have a circular Gaussian distribution. The transmitted field in output mode $b$ located in the image plane is
\begin{equation}
E_b =\sum_{a=1}^Nt_{ba}E_a,
\label{eq:transmission}
\end{equation}
where $E_a$ is the input field coming from the phase modulator. We optimize the incident field, $E_a$, by getting feedback from the intensity of multiple targets located in the back focal plane. The field in the back focal plane is the Fourier transform of the field in the image plane \cite{goodman2005introduction}. The field in target $m$ placed in the back focal plane is
\begin{equation}
E^m_k \propto \sum_{b}E_b \exp(-ik_mb),
\label{eq:Fourier}
\end{equation}
where $k_m$ is the wave vector component. After maximizing the intensity of $M$ targets in the back focal plane, the optimized incident field of input mode a, $\hat{E}{a}$, is decomposed into a contribution of the ideal incident field and an orthogonal part to the ideal field, $\zeta^m_a$, accounting for experimental imperfections. $\hat{E}{a}$ is given by
\begin{equation}
{\hat{E}}_{a}=D\gamma\sum_{m=1}^{M}\sum_{b}t^*_{ba}\exp(ik_mb) + \sum_{m=1}^{M}\sqrt{\frac{1-|\gamma|^2}{NM}}\zeta^m_a
\label{eq:Ea}
\end{equation}
where the prefactor $D=(\,NM\sum_{b} \langle |t_{ba}|^2 \rangle)\,^{-1/2}$ normalizes the total incident intensity by assuming that the fields for different $M$ targets and $N$ SLM segments are orthogonal. $\zeta^m_a$ has a complex Gaussian distribution with mean zero and standard deviation one. $^*$ represents the complex conjugate. The fidelity parameter, $|\gamma|^2$, for phase only modulation is given by \cite{yilmaz2013optimal}
\begin{equation}
|\gamma|^2= \frac{\pi}{4}\frac{SNR}{1+SNR} ,
\label{eq:fidelity}
\end{equation}
where $SNR$ is the signal to noise ratio of the feedback signal used for wavefront shaping. Replacing the incident field in Eq. (\ref{eq:transmission}) by Eq. (\ref{eq:Ea}) gives 
\begin{equation}
\hat{E}{b} = \sum_{a=1}^Nt_{ba} D\gamma \sum_{m=1}^{M}\sum_{b'}t^*_{b'a}\exp(ik_{m}b') + \sum_{a=1}^Nt_{ba}\sum_{m=1}^{M}\sqrt{\frac{1-|\gamma|^2}{NM}}\zeta^m_a,
\label{eq:Eb^_c}
\end{equation}
which is the field in the image plane corresponding to $M$ optimized targets in the back focal plane. The ensemble averaged optimized intensity in the image plane is
\begin{equation}
\langle|\hat{E}_{b}|^2\rangle = |\gamma|^2M(N-1)\frac{\langle|t_{ba}|^2\rangle^2}{\sum_{b'}\langle |t_{b'a}|^2\rangle}+\langle |t_{ba}|^2\rangle,
\label{eq:I_beta}
\end{equation}
where the angle brackets denote ensemble averaging over disorder. 
The enhancement of the constructed focus in output mode $\beta$, $\eta(\beta)$, is defined as the ratio of the optimized intensity (Eq. \ref{eq:I_beta}) to the ensemble averaged intensity, $\langle|{E}_{b}|^2\rangle = \langle|t_{ba}|^2\rangle$.
\begin{equation}
\eta(\beta) = |\gamma|^2M(N-1) \frac{\langle|t_{\beta a}|^2\rangle}{\sum_{b} \langle |t_{ba}|^2 \rangle}+1
\label{eq:etab}
\end{equation}

We define the number of speckles as $N_s\equiv {\sum_{b}I_{b}}/{\max I_b},$ with $I_{b}=\langle|t_{ba}|^2\rangle$, the ensemble averaged diffuse intensity. Rewriting Eq. (\ref{eq:etab}) gives 

\begin{equation}
\eta(\beta) = \frac{|\gamma|^2M(N-1)}{N_s}\frac{\langle|t_{\beta a}|^2\rangle}{\max{\langle |t_{ba}|^2\rangle}}+1
\label{eq:eta2}
\end{equation}

When we substitute $F(\beta) = \langle|t_{\beta a}|^2\rangle/{\max{\langle |t_{ba}|^2\rangle}}$ in Eq. (\ref{eq:eta2}), we derive Eq. (\ref{eq:enhancement}) describing that the enhancement follows the distribution of the original diffuse intensity.

\section*{Funding}
This work was financially supported by the Nederlandse Organisatie voor Wetenschappelijk Onderzoek (TTW-NWO, Vidi grant 14879).
\section*{ACKNOWLEDGEMENTS}
We would like to thank Gerwin Osnabrugge for the helpful discussions.

\bibliography{sample.bib}

\end{document}